\documentclass[aps,pra, twocolumn, twoside,floatfix,superscriptaddress,showpacs]{revtex4}
\usepackage{hyperref}
\usepackage{bbm}
\usepackage{amsmath, amssymb}
\usepackage{color}
\usepackage{graphicx,epsfig}
\usepackage{pifont} 
\usepackage{bbold}

\def\half{\frac{1}{2}}

\newcommand{\red}{\color{black}}  
\newcommand{\mg}{\color{black}} 
\newcommand{\blk}{\color{black}}
\newcommand{\blu}{\color{black}} 
\definecolor{maroon}{rgb}{0.7,0,0}

\definecolor{ngreen}{rgb}{0.2,0.6,0.2}

\definecolor{golden}{rgb}{0.8,0.6,0.1}

\begin{document}

\title{Canonical form of master equations and characterization of non-Markovianity}

\author{Michael J. W. Hall}
\affiliation{Centre for Quantum Computation and Communication Technology (Australian Research Council), Centre for Quantum Dynamics, Griffith University, Brisbane, QLD 4111, Australia}
\author{James D. Cresser}
\affiliation{Department of Physics and Astronomy, Macquarie University, Sydney NSW 2109, Australia}
\author{Li Li}
\affiliation{Centre for Quantum Computation and Communication Technology (Australian Research Council), Centre for Quantum Dynamics, Griffith University, Brisbane, QLD 4111, Australia}
\author{Erika Andersson}
\affiliation{SUPA, Institute for Photonics and Quantum Sciences, School of Engineering and Physical Sciences, Heriot-Watt University,
Edinburgh, EH14 4AS, UK}

\begin{abstract}
Master equations govern the time evolution of a quantum system interacting with an environment, and may be written in a variety of forms. Time-independent or memoryless master equations, in particular, can be cast in the well-known Lindblad form.  
Any time-local master equation, Markovian or non-Markovian, may in fact also be written in a Lindblad-like form. A diagonalisation procedure results in a unique, and in this sense canonical, representation of the equation, which may be used to  fully 
characterize the non-Markovianity of the time evolution.
Recently, several different measures of non-Markovianity have been presented which reflect, to varying degrees, the appearance of negative decoherence rates in the Lindblad-like form of the master equation. We therefore propose using the negative decoherence rates themselves, as they appear in the canonical form of the master equation, to completely characterize non-Markovianity. The advantages of this are especially apparent when more than one decoherence channel is present.  We show that a measure proposed by Rivas {\it et al.} is a surprisingly simple function of the canonical decoherence rates, and give an example of a master equation that is non-Markovian for all times $t>0$, but to which nearly all proposed measures are blind. We also give necessary and sufficient conditions for trace distance and volume measures to witness non-Markovianity, in terms of the Bloch damping matrix.
\end{abstract}

\maketitle

\section{Introduction}

An open quantum system is a quantum system whose dynamics is determined both by interactions internal to the system, and by influences from an environment. As no physical system is truly isolated, this is a situation which applies very widely. Time-independent or memoryless behavior leads to Markovian master equations in the so-called Lindblad form~\cite{Lindblad, Gorini}.
Lindblad master equations have been extensively used to describe phenomena in  e.g. quantum optics, semiconductor physics and atomic physics, ranging from the decay of an atom to a quantum-mechanical description of Brownian motion~\cite{BreuerPetroBook, friction,wiseman}.

The non-Markovian case is less well understood, but is becoming increasingly relevant as our ability to experimentally control quantum systems develops. Simple examples are a damped harmonic oscillator and a damped driven two-level atom~\cite{BreuerPetroBook,wiseman}.
Also, the vast majority of calculations of error thresholds for quantum computing makes the assumption that the noise processes are Markovian. This is not necessarily physically realistic~\cite{Lidar}.

Recently, several different measures of non-Markov\-ianity have been proposed~\cite{Wolf, Breuermeasure, Rivas, Lu,neg, inf, distance, bures,volume}. For example, the measure in \cite{Breuermeasure} is based on whether it is possible for the trace distance between two initial states to increase as a function of time. If the time evolution starting from a time $t_0$ is completely positive, then the trace distance cannot exceed its inital value at $t_0$, but it could first decrease and then increase again. This would be a signature of non-Markovian behavior. Other signatures of this type are based on increases in entanglement \cite{Rivas,neg,inf}, Fisher information and Bures distance \cite{Lu,distance,bures}, and the volume of states \cite{volume}. Measures directly based on the definition of complete positivity, such as the amount of isotropic noise needed to make the evolution completely positive \cite{Wolf}, have also been suggested \cite{Wolf,Rivas}.

The above measures are all compatible with a definition of Markovianity corresponding to being able to write the master equation in a local-in-time Lindblad-like form, such that all decoherence rates $\gamma_k(t)$ are positive at all times. Mathematically equivalent characterizations for finite-dimensional systems are that the evolution is divisible into a sequence of infinitesimal completely positive evolutions \mg \cite{Rivas,Wolf2} \blk (analogous to the classical Chapman-Kolmogorov equation \cite{gardiner}), and that the evolution can be modelled by standard quantum jump trajectories \mg for the system \blk \cite{breuer4, Piilo1, Piilo2}.
Hence, a natural definition of {\it non}-Markovian time evolution is the unavoidable appearance of one or more negative decoherence rates in the master equation. 

In order to obtain a fundamental 
characterization underlying the many proposed measures of non-Markovianity,
 it follows that one should focus on their
common point of origin, i.e., the negative decoherence rates $\gamma_k(t)$ themselves. We will show that this leads to a unique measure that gives a complete picture of non-Markovianity even when there are several decoherence channels present. Also, it does not require solving the master equation to obtain the time evolution of the system, or optimising over initial states.

Any function of the $\gamma_k(t)$ that can witness their negativity in at least some cases will be a valid measure of non-Markovianity, and the measures in~\cite{Wolf, Breuermeasure, Rivas, Lu,neg, inf, distance, volume}
can be viewed in this way.
Corresponding signatures, such as increasing trace distance or increasing entanglement, are useful as experimentally accessible indicators of the presence of negative decoherence rates \cite{exp1,exp2}. If only one decoherence channel is present, these measures are more or less equivalent to knowledge of 
$\gamma(t)$~\cite{Breuermeasure, Rivas}.  However, more generally, they may not witness non-Markovian evolution even when it is present.

If one is to base a measure of non-Markovianity directly on the negativity of decoherence rates in a master equation, it is essential to have a unique form of the equation, since each master equation may be written in many ways. This is indeed a reason cited in~\cite{Breuermeasure} for not basing a definition of non-Markovanity directly on a master equation. Fortunately, there is such a unique and canonical form. This result is a straightforward extension of the treatment in \cite{Gorini}. General forms of time-local master equations have been used previously e.g. in~\cite{Wolf2,vanwond,breuer4}, but the significance and usefulness of the unique diagonal form of the equation has, as far as we are aware, not been emphasized. 
{Neither is the generality of time-local master equations~\cite{DanielPRA2012} as widely recognised as it should be.} We will therefore start by presenting  the canonical form of a time-local master equation in Sec.~II, before discussing how this gives rise to a natural means of completely quantifying non-Markovianity in Sec.~III, in terms of the canonical decoherence rates.

In Sec.~IV we consider various proposed measures of non-Markovianity \cite{Wolf, Breuermeasure, Rivas, Lu,neg, inf, distance, volume}, and assess their relative abilities to witness non-Markovianity.  We find, surprisingly, that none of the measures based on an increasing distance, volume or entanglement can witness the non-Markovianity of a simple qubit example.  However, the Choi-matrix based measures of Wolf {\it et al.} \cite{Wolf} and Rivas {\it et al.} \cite{Rivas} are always faithful witnesses of non-Markovianity, and are shown to be simple functions of the canonical decoherence rates. We also give simple necessary and sufficient conditions for trace distance and volume measures to be able to witness non-Markovianity, in terms of the Bloch damping matrix.

Conclusions are presented in Sec. V. Technical details are largely deferred to the Appendices.

\section{Forms of master equations}

\subsection{Memoryless master equations}

Under fairly general conditions, a master equation for a system density operator $\rho$ takes the form \cite{Naka, 20} 
\begin{equation}
\label{genmaster}
\dot{\rho}(t)=-\frac{\rm i}{\hbar}\left[H_S,\rho(t)\right]+\int_{0}^{t}\mathcal{K}_{s,t}[\rho(s)]\,ds.
\end{equation}
Here $H_S$ is the system Hamiltonian, 
and the memory kernel $\mathcal{K}_{s,t}$ is a linear map describing the effects of the environment on
the system.  The Born-Markov approximation amounts to approximating the memory kernel \red (in the interaction picture) \blk
by \red
$\mathcal{K}_{s,t}\approx \delta(t-s)\,\mathcal{K}_t$.  Further, any explicit time dependence in $\mathcal{K}_t$ typically comprises rapidly fluctuating terms, that may be removed via a secular (or random wave) approximation to give a memoryless master equation of Lindblad form \cite{BreuerPetroBook}:
\begin{equation}
\label{lindbladeqn}
\dot{\rho}
= -\frac{\rm i}{\hbar}\left[H,\rho\right]+ \sum_k \gamma_k 
\left(\hat{L}_k \rho \hat{L}_k^\dagger-\frac{1}{2}\left\{ \hat{L}_k^\dagger \hat{L}_k ,\rho\right\}  \right) .
\end{equation}
Here the $\gamma_k$ are positive constants, called decoherence rates, and the unitary part of the time evolution corresponding to $H$ may include effects arising from the environment \cite{BreuerPetroBook}. The Lindblad form in Eq.~(\ref{lindbladeqn}) may also be derived under the assumptions that the evolution is completely positive and generated by a dynamical semigroup~\cite{Lindblad}.  
\blk
 
\subsection{Time-local master equations}

More generally, even when the system `remembers' its previous evolution, it has been shown via the time-convolutionless projection operator method \cite{Chat,Shib}
that a general master equation (\ref{genmaster}) often can be written in a time-local form,
$	\dot{\rho}(t)=\Lambda_{t}[\rho(t)]$,
where $\Lambda_{t}$ is, among other requirements, 
a linear map such that $\Lambda_{t}[\rho]$ is Hermitian and traceless for all $\rho$.  \blu An alternative approach, for the class of master equations obtained from the reduced memoryless evolution of a system plus ancilla, is given in \cite{koss}.  \blk
Such a description of the evolution is not only convenient, but appears necessary, for example, to construct a quantum trajectory unravelling in terms of possible `paths' the time evolution of the system may take~\cite{Piilo1,carmichael, plenio, Breuer, breuer4}.

For an alternate and very simple proof that memory-kernel master equations can typically be written in time-local form~\cite{ourmaster},
consider some evolution process described by the linear (usually completely positive) map $\rho(t) =
\phi_t[\rho(0)]$, which satisfies a memory-kernel master equation of
the general form in Eq.~\eqref{genmaster}. If it is assumed that the map $\phi_t$ is {\it
invertible} for the time interval considered, i.e., that there exists a linear map $\phi^{-1}_t$
satisfying $\phi^{-1}_t\circ \phi_t=\mathbbm{1}$ where $\mathbbm{1}$ is the identity map, then one can write,  absorbing the Hamiltonian component into $\mathcal{K}_{s,t}$,
\begin{eqnarray}
\dot{\rho}(t) &=& \int_0^t ds\, (\mathcal{K}_{s,t}\circ\phi_s)[\rho(0)]\nonumber\\
&=& \int_0^t ds\,
(\mathcal{K}_{s,t}\circ\phi_s\circ\phi^{-1}_t)\{\phi_t[\rho(0)]\}\nonumber\\
\label{simple} 
&=& \Lambda_t[\rho(t)],
\end{eqnarray}
where $\Lambda_t:= \int_0^t ds\,
\mathcal{K}_{s,t}\circ\phi_s\circ\phi^{-1}_t$, which is of time-local (albeit in general complicated)
form.

The assumption that the evolution is invertible is not strong, and is violated only if two initially distinct states evolve to the same state
at some finite time $t$ (e.g., if an `equilibrium state' is reached within a finite time rather than asymptotically).  As shown in~\cite{ourmaster}, even in this case it is sometimes possible to describe the evolution by a
time-local master equation. Typically, decoherence rates approach infinity at times when the evolution is not invertible~ \blu \cite{ourmaster, diosi, jimheis,koss}. \blk Note also that even if the time evolution is not invertible for isolated points in time, but a time-local master equation exists for other times, it can still be used for a characterization of non-Markovianity.
Physically relevant examples where the time evolution is invertible for all times include spontaneous emission and optical phase diffusion \cite{Jimphase}, where the latter is also an example of a time-local master equation corresponding to non-Markovian evolution.

\subsection{Canonical form for master equation}

The Lindblad form for memoryless master equations in Eq.~(\ref{lindbladeqn}) is non-unique, as $\gamma_k$ and $L_k$ can be chosen in infinitely many different ways, corresponding to different Kraus decompositions of completely positive maps \cite{Lindblad}. 
Nevertheless, as shown in Appendix~A by explicit construction, any time-local master equation, $\dot \rho=\Lambda_{t}[\rho(t)]=\sum_k A_k(t)\rho B_k(t)$, can be cast in a canonical Lindblad-like form that uniquely defines a set of corresponding `canonical' decoherence rates.  Hence, this form is suitable for characterizing non-Markovianity. 
 
The key to obtaining the canonical form  is to first rewrite the master equation in terms of a corresponding time-dependent `decoherence matrix'.  The eigenvalues and eigenvectors of this matrix directly determine the canonical decoherence rates $\gamma_k(t)$ and decoherence operators $L_k(t)$,  and the corresponding canonical decoherence channels are mutually orthogonal in the sense defined below. The derivation is a straightforward extension of the approach in~\cite{Gorini}  (see \cite{Wolf2,eprint}), and deserves to be more generally known. 

In particular, as shown in 
Appendix~A, any local-in-time master equation, for a quantum system having a $d$-dimensional Hilbert space, can be written in the {\it canonical} form
\begin{eqnarray}	\nonumber
	\dot{\rho}
	&=&-\frac{i}{\hbar} [H(t),\rho] 
	+\sum_{k=1}^{d^2-1}\gamma_k(t)\left[L_k(t)\rho L_k^\dagger(t) \right. \\
&~&~~~~~~~~~~~~~~~~~~~~~~~~~~~~~~\left. -\half \left\{ L_k^\dagger(t) L_k(t),\rho \right\} \right] \label{canonical}
\end{eqnarray}
where  the $L_k(t)$ form an orthonormal basis set of traceless operators, i.e.,
\begin{equation} \label{orthog}
\text{Tr}[L_k(t)] = 0,~~~~~\text{Tr}[L_j^\dagger(t)\, L_k(t)] =\delta_{jk},
\end{equation}
and $H(t)$ is Hermitian \cite{trace}.

Equation~(\ref{canonical}) may be recognised as being similar to the Lindblad form for memoryless master equations in Eq.~(\ref{lindbladeqn}).  However, unlike Eq.~(\ref{lindbladeqn}): 
\begin{description}
\item[(i)]{ the decoherence rates $\gamma_k(t)$ and the decoherence operators $L_k(t)$ are in general time dependent;} 
\item[(ii)]{ the decoherence rates $\gamma_k(t)$ are uniquely determined, and, moreover, remain invariant under any  unitary transformation $\rho\rightarrow V(t)\rho V(t)^\dagger$ (e.g. to an `interaction' picture);}
\item[(iii)]{ the decoherence rates can be negative, corresponding to interactions between the environment and the system in such a way that  the system may {\it recohere}, reversing earlier decay \mg processes~\cite{Piilo1, Breuer, maniscalcomisbelief, ourmaster}; \blk and}
\item[(iv)]{ the decoherence operators $L_k(t)$ are restricted to correspond to a set of `orthogonal' decoherence channels, as per Eq.~(\ref{orthog}).}
\end{description}

Note that constraints on  $\gamma_k(t)$ and $L_k(t)$ to ensure that the time evolution is completely positive, or even positive,
have not been considered here. Thus, for instance, while all time-local master equations can be written in the above canonical form, there is no guarantee that an arbitrary master equation of this form will yield only positive eigenvalues for $\rho$ for all times.  Conditions for time-local qubit master equations to generate completely positive evolution are discussed in \cite{hall08}.

\section{Characterizing non-Markovianity}
 
\subsection{Definition}

It is seen that time-local master equations are widely applicable, and that any such master equation may be rewritten in the canonical form \eqref{canonical}.  Further, this form uniquely determines a set of corresponding canonical decoherence rates, $\gamma_k(t)$, given by the eigenvalues of the decoherence matrix $\mathsf{\mathbf{d}}$ in Eq.~(\ref{dexplicit}). An alternative means for determining the $\gamma_k(t)$ is given in Appendix C.

If the canonical decoherence rates are positive at all times, then the evolution over any time interval is completely positive.  This has been termed `time-dependent Markovian' evolution, e.g., in~\cite{Wolf, Breuermeasure, Rivas}, and clearly generalises the completely positive evolution guaranteed  for memoryless master equations in Eq.~(\ref{lindbladeqn}). Moreover, for finite systems, having positive decoherence rates for all times is equivalent to  the divisibility of the evolution into a sequence of infinitesimal completely positive evolutions \mg \cite{Rivas, Wolf2}; \blk and to being able to model the evolution by standard quantum jump trajectories for the system \cite{breuer4, Piilo1, Piilo2}.  

In light of the above, and as foreshadowed in the introduction, we are led to the following.
\begin{description}
\item[Definition]{: A time-local master equation is {\it Markovian}, at some given time, if and only if the canonical decoherence rates are positive.  Correspondingly, the evolution is {\it non-Markovian} if one or more of the canonical decoherence rates is strictly negative.}
\end{description}

To illustrate the importance of using the canonical form of the master equation to characterize non-Markovianity,
consider the evolution corresponding to
\begin{eqnarray*}
\dot{\rho}&=&[2\gamma(t)+\tilde{\gamma}(t)]\left[2\sigma_x\rho\sigma_x +2\sigma_y\rho\sigma_y 
-4\rho\right]\nonumber\\
&&-\gamma(t)\left[2\sigma_-\rho\sigma_+ -\sigma_+
\sigma_-\rho-\rho\sigma_+ \sigma_- \right]\\
&&-\gamma(t)\left[2\sigma_+\rho\sigma_- -\sigma_-
\sigma_+\rho-\rho\sigma_- \sigma_+ \right].
\end{eqnarray*}
At first sight this may appear to be non-Markovian whenever $\gamma(t)> 0$ {\it or} $2\gamma(t)+\tilde{\gamma}(t)<0$.  However, 
the canonical form of this equation may be written as
\[
\dot{\rho}=[\gamma(t)+\tilde{\gamma}(t)]\left[2\sigma_x\rho\sigma_x +2\sigma_y\rho\sigma_y 
-4\rho\right],
\]
and hence the evolution is non-Markovian if and only if $\gamma(t) +\tilde{\gamma}(t)< 0$.

\blu
A more general example showing the necessity of using the canonical form is the master equation
\begin{eqnarray*} 
\dot \rho &=& L\rho L^\dagger - \half (L^\dagger L\rho +\rho L^\dagger L)\\
&~& - \big[ L^\dagger \rho L - \half (L L^\dagger \rho +\rho L L^\dagger ) \big] ,
\end{eqnarray*}
where the second line appears, {\it prima facie}, to generate non-Markovian evolution for any nonzero $L$.  However, making the choice $L=(1+iH/\hbar)/\sqrt{2}$, for any Hermitian operator $H$, this master equation reduces to the Hamiltonian form
\[ \dot \rho = -\frac{i}{\hbar} [H,\rho], \]
corresponding to unitary (and hence trivially Markovian) evolution.
\blk

Finally, it is worth emphasising again that the canonical decoherence rates, and hence the above characterization of non-Markovianity, are not only unique, but are invariant under any time-dependent unitary transformation of the system (see previous section and Appendix~A).  

\subsection{Canonical measures of non-Markovianity}

Since the defining feature of non-Markovianity is whether {any of} the $\gamma_k(t)$ in the canonical form  (\ref{canonical}) become negative, it is natural to make use of the functions 
\begin{equation} \label{fk}
f_k(t):=\max[0,-\gamma_k(t)] \, \geq 0
\end{equation}
to describe the non-Markovanity in individual decoherence channels. 
We can either use the $f_k(t)$ directly, as canonical measures of non-Markovianity at time $t$, or  a function such as 
\begin{equation} \label{ft}
f(t)= \sum_{k=1}^{d^2-1} f_k(t) =\half\sum_{k=1}^{d^2-1} \big[|\gamma_k(t)|-\gamma_k(t)\big]
\end{equation}
\mg (rescaled \blk by some function of $d$ if we wish). 

We can similarly use the corresponding integrals
\begin{equation}
F_k(t,t')=  \int_t^{t'} ds\, f_k(s)  
\end{equation} 
to characterize the ``total amount of non-Markovianity in channel $k$ over the time interval $[t,t']$'', and the corresponding sum
\begin{equation} \label{total}
F(t,t')=\sum_{k=1}^{d^2-1} F_k(t,t') =  \int_t^{t'} ds\, f(s) 
\end{equation}
as a measure of the ``total amount of non-Markovianity over the interval $[t,t']$''. Either of these quantities may be \mg rescaled \blk -- e.g., by the total elapsed time $t'-t$. Note that $f(t)$ and $F(t,t')$ are strictly positive if and only if the evolution is non-Markovian.  It will be shown in Sec.~IV that a measure proposed by Rivas {\it et al.} \cite{Rivas} is equal to $2d^{-1}F(0,\infty)$, thereby providing a simple formula (and interpretation) for the proposed measure, in terms of the canonical decoherence rates.

We also define a canonical discrete measure of non-Markovianity, as the number of strictly negative decoherence rates, i.e.,
\begin{equation} \label{index}
n(t) :=\, \#\{k:\gamma_k(t) <0\} =\, \#\{ k: f_k(t)>0\} .
\end{equation}
Thus, $n(t)$ is a ``non-Markov index''. 

In the sense that ${\rm Tr}[L_j(t)^\dagger L_k(t)]=\delta_{jk}$, as per Eq.~(\ref{orthog}), the non-Markovian part of the dynamics takes part in a region of `evolution space' which is orthogonal to the Markovian region. For example, for a two-level system, if Markovian behavior is generated by the $x$ direction of the Bloch vector, then non-Markovian behavior can be generated by the $y$ and $z$ directions. The non-Markov index therefore characterizes the dimension of the space of non-Markovian evolution.

\subsection{Example: single decoherence channel}

The simplest case to consider is a master equation with only one non-zero decoherence rate, i.e., of the form
$ \dot \rho = -\frac{i}{\hbar} [K(t),\rho] + \alpha(t) \left[ A(t)\rho A(t)^\dagger -\frac{1}{2} \{A(t)^\dagger A(t),\rho\}\right]. $
The corresponding canonical form is easily checked to be
\begin{equation} 
\dot \rho = -\frac{i}{\hbar} [H(t),\rho] + \gamma(t) \left( L(t)\rho L(t)^\dagger -\frac{1}{2} \{L(t)^\dagger L(t),\rho\}\right), 
\end{equation}
with $L(t):=(A-a)/\{\text{Tr}[(A^\dagger-a^*) (A-a)] \}^{1/2}$,  $\gamma(t) :=\alpha \text{Tr}[(A^\dagger-a^*) (A-a)]$, 
$H(t):=K -\frac{1}{2} i\hbar \alpha\left[ a\,A^\dagger - a^*A\right]$,
and $a(t):=d^{-1}\text{Tr}[A(t)]$. These definitions ensure that that $\text{Tr}[L(t)]=0$ and $\text{Tr}[L(t)^\dagger L(t)]=1$, as required by Eq.~(\ref{orthog}).

Hence, from the definition of non-Markovianity in Sec.~III~A, the evolution of a system with a single decoherence channel is non-Markovian, at time $t$, if and only if $\gamma(t)<0$ (or,  equivalently, $\alpha(t)<0)$. The total amount of non-Markovianity over an interval $[t,t']$ follows from Eq.~(\ref{total}) as 
\begin{equation} \label{fex}
 F(t,t') =- \int_{\gamma(t)<0} ds\,\gamma(s) .
\end{equation}
Further, from Eq.~(\ref{index}), the non-Markov index is unity when $\gamma(t)<0$ and zero otherwise.

The case of a single decoherence channel appears to be the prototypical example used in the literature to assess various proposed measures of non-Markovianity \cite{Wolf, Breuermeasure, Rivas, Lu,neg, inf, distance, volume}.  
However, this case is too simple to assess the relative strengths of the proposed measures, as they  are all functionals of $\gamma(t)$ that are  sensitive to its sign. For example,  the rate of change of  trace distance for the qubit example in \cite{Breuermeasure} is $-\gamma(t)\exp[-\int_0^t ds \,\gamma(s)]$, while the Choi-matrix based measure ${\cal I}$ in \cite{Rivas} evaluates to $F(0,\infty)$ with $F$ as in Eq.~(\ref{fex}). 


For this reason, examples of master equations with {\it multiple} decoherence channels are required to properly assess and compare proposed measures of non-Markovianity (see also Sec.~IV).

\subsection{Example: eternal non-Markovianity}

The qubit master equation 
\begin{equation} \label{ex}
\dot \rho = \frac{1}{2}\sum_{k=1}^3\gamma_k(t) \left[\sigma_k \rho\sigma_k -\rho\right],
\end{equation}
where the $\sigma_k$ are the Pauli sigma matrices and
\begin{equation} \label{exgamma}
\gamma_1(t)=\gamma_2(t)=1,~~~\gamma_3(t)=-\tanh t,
\end{equation}
is of particular interest, as it provides a simple example of a completely positive evolution that is non-Markovian at all times $t>0$, yet which is not detectable as non-Markovian by the majority of proposed measures in the literature (see Sec.~IV).  

Complete positivity of the map from $\rho(0)$ to $\rho(t)$, for a master equation of the form in Eq.~(\ref{ex}), corresponds to  $\Gamma_j+\Gamma_k\leq 1+\Gamma_l$ for all permutations $j,k,l$ of $1,2,3$, with $\Gamma_j(t):=\exp\{-\int_0^tds\,[\gamma_k(s)+\gamma_l(s)]\}$ \cite{ourmaster,hall08}, which is always satisfied by the above example.  Such master equations are straightforward to solve, and in terms of the Bloch vector  ${\bf x}$, with $x_k:=\text{Tr}[\rho {\sigma_k}]$ and $\rho=\frac{1}{2}[1+\sigma.{\bf x}]$, one finds
\[ x_j(t)=\half(1+e^{-2t})x_j(0)~(j=1,2),~x_3(t)=e^{-2t}x_3(0) .
\]
Thus, the initial Bloch vector asymptotically evolves to its projection on the $xy$-plane, scaled by a factor of $\half$. 

Equation~(\ref{ex}) is in canonical form (identifying $L_k(t)=\sigma_k/\sqrt{2}$), and thus the system has three orthogonal decoherence channels.  It is clearly non-Markovian for all $t>0$, so that the third channel could even be termed a `recoherence' channel. The canonical measure of total non-Markovianity over the interval $[0,t]$ follows from Eq.~({\ref{total}) as
\begin{equation}
F(0,t) = -\int_0^t ds\,\tanh s = \ln \cosh t .
\end{equation}
Thus, the average non-Markovianity, $F(0,t)/t$, approaches unity as $t\rightarrow\infty$.  The non-Markov index follows from Eq.~(\ref{index}) as $n(t)=1$ for all $t>0$, corresponding to non-Markovian behaviour with respect to the $z$-direction.  Even so, this `eternal' non-Markovianity cannot be detected by various distance, volume and entanglement measures, as shown in the following section.

\section{Relative strengths of different measures of non-Markovianity}

Previous measures of non-Markovianity in the literature are based on the relationship between Markovian evolution and completely-positive maps \cite{Wolf, Breuermeasure, Rivas, Lu,neg, inf, distance, bures, volume}.  In particular, for Markovian evolution at time $t$, the decoherence rates in Eq.~(\ref{canonical}) are all positive, by definition, and hence the infinitesimal map taking the system from time $t$ to time $t+dt$ must be completely positive \cite{Wolf2}.  It follows, for example, that any quantity that decreases under completely-positive maps provides a suitable signature of non-Markovianity: if this quantity is found to {\it increase} at time $t$, then the evolution must be non-Markovian.

Proposed signatures of the above type leads to measures of non-Markovianity that fall into three broad categories -- distance based measures, volume based measures, and entanglement based measures.  A fourth category of interest is based on the Choi-matrix representation of completely positive maps. These categories are examined in turn below.  Of the measures examined, only those based on the Choi matrix can detect the non-Markovianity of the example in Eqs.~(\ref{ex}) and (\ref{exgamma}).

\subsection{Distance measures}

\subsubsection{Trace distance}

Breuer {\it et al.} have suggested using the increasing of trace distance between two states as a signature of non-Markovianity \cite{Breuermeasure}, and this signature has subsequently been experimentally investigated for qubits \cite{exp1,exp2}.  

In the Bloch representation, the square of the trace distance between any two infinitesimally separated qubit density operators, $\rho$ and $\rho+\delta\rho$, is
\begin{equation} \label{tracedist}
(\delta s_{\rm Tr})^2 :=  \frac{1}{4} \left(\text{Tr}|\delta\rho|\right)^2 = \frac{1}{4} \delta{\bf x}\cdot \delta{\bf x},
\end{equation}
corresponding to a Euclidean metric on the space of Bloch vectors.  For qubits, it is well known that any master equation can be rewritten in the Bloch representation as 
\begin{equation} \label{bloch}
\dot{\bf x} = D(t){\bf x} + {\bf u}(t), 
\end{equation}
where ${\bf x}$ denotes the Bloch vector, and $D$ and ${\bf u}$ are referred to as the damping matrix and drift vector, respectively. Hence, $\delta \dot{\bf x}=D\delta{\bf x}$, and so
\[
\frac{d}{dt} (\delta s_{\rm Tr})^2 = \frac{1}{4}\left[\delta\dot{\bf x}\cdot \delta{\bf x} + \delta{\bf x} \cdot \delta\dot{\bf x}\right] = \frac{1}{4} \delta{\bf x}^T (D+D^T)\delta{\bf x},
\] 
where the superscript $T$ denotes the transpose.  

Now, the trace distance can increase between some pair of density operators if and only if it can increase between some pair of infinitesimally separated density operators.  But the above equation shows that latter  is possible if and only if the matrix $D+D^T$ has a positive eigenvalue.  That is, {\it the qubit trace distance can witness non-Markovianity at time $t$ if and only if the damping matrix  satisfies the condition}
\begin{equation} \label{tracecond}
\lambda_{\rm max}\left[D(t)+D^T(t)\right] > 0 
\end{equation}
where $\lambda_{\rm max}(A)$ denotes the maximum eigenvalue of $A$. This result subsumes the previously-considered special case of random unitary evolution \cite{random}.

For the case of a single decoherence channel the above condition is equivalent to $\gamma(t)<0$, and hence non-Markovianity can always be witnessed for such channels.  However, for the qubit example in Eqs.~(\ref{ex}) and (\ref{exgamma}), the damping matrix may be calculated as \cite{hall08}
\begin{equation} \label{damp}
 D(t) = \left(
\begin{array}{ccc}
-1+\tanh t&0&0\\
0&-1+\tanh t&0\\
0&0&-2
\end{array}
\right) \leq 0 .
\end{equation}
Hence, the trace distance can never witness the non-Markovianity of this example. 

\subsubsection{Bures distance and Fisher information}

Other measures of distance which decrease under completely positive maps, such as the Bures and Hellinger distance, have similarly been suggested as signatures of non-Markovianity. \cite{Lu,distance,bures}.  Such distances are typically generated by some monotone metric $G\equiv g_{jk}$ on the space of density operators \cite{monotone}, and for qubits have the infinitestimal form
\begin{equation} \label{metric}
(\delta s)^2 = g_{jk}({\bf x}) \delta x_j\delta x_k = \delta{\bf x}^T \mg G({\bf x}) \blk \,\delta{\bf x}
\end{equation}
for Bloch vectors ${\bf x}$ and ${\bf x}+\delta{\bf x}$ (the higher-dimensional case is similar, but will not be explicitly considered here). For example, the qubit Bures distance corresponds to the (maximally symmetric) quantum Fisher information metric \cite{hubner}
\begin{equation} \label{bures}
4(\delta s_B)^2 = \frac{({\bf x}\cdot \delta{\bf x})^2}{1-{\bf x}\cdot {\bf x}} + \delta {\bf x}\cdot \delta{\bf x}.
\end{equation}
Note while attention can be restricted to families of density operators defined by some real parameter, $\theta$ or $\lambda$ say \cite{Lu, distance}, we consider arbitrary density operators here, parameterised by the Bloch vector, for full generality.

Just as for the trace distance, a given distance measure can witness non-Markovianity at time $t$ if and only if the distance between some pair of infinitesimally separated density operators is increasing.  For qubits, using Eqs.~(\ref{bloch}) and (\ref{metric}), this corresponds to the existence of Bloch vectors ${\bf x}$ and ${\bf x}+\delta{\bf x}$ such that
\[
\frac{d}{dt} (\delta s)^2 = \delta{\bf x}^T \left[ \dot G + GD+D^TG\right] \delta{\bf x} > 0 .
\]
This is equivalent to to the existence of a positive eigenvalue of the symmetric matrix $J:=\dot G + GD+D^TG$, i.e., to  
\begin{equation}
\lambda_{\rm max}\left[\dot G + GD+D^TG\right] > 0 
\end{equation}
at time $t$, which clearly generalises Eq.~(\ref{tracecond}) for the trace distance.

For the case of the Bures distance in Eq.~(\ref{bures}), and a qubit master equation of the form of Eq.~(\ref{ex}), the corresponding
 matrix $J_B$ follows via differentation of Eq.~(\ref{bures}) with respect to time. Using Eq.~(\ref{bloch}) with ${\bf u}(t)={\bf 0}$, this yields
\begin{equation} \label{vb}
4J_B = \bar D +\frac{\bar D {\bf x}{\bf x}^T +  {\bf x}{\bf x}^T\bar D}{1-{\bf x}\cdot{\bf x}} + \frac{({\bf x}^T\bar D{\bf x})\, {\bf x}{\bf x}^T}{(1-{\bf x}\cdot{\bf x})^2}  ,
\end{equation}
where $\bar D:=D+D^T$.
We have checked numerically, for $D$ as in  Eq.~(\ref{damp}), that $J_B$ is negative definite for all times $t>0$.  Hence, just as for the trace distance, the Bures distance cannot witness the non-Markovianity of the example in Eqs.~(\ref{ex}) and (\ref{exgamma}).  It remains an open question as to whether some other distance measure can do so.

\subsection{Bloch volume measure}

Lorenzo {\it et al.} have proposed using the increase of volume of the set of states of a system as a signature of non-Markovianity \cite{volume}.  Here, the volume measure corresponds to the determinant of the evolution map in the (generalized) Bloch representation of the system, which cannot decrease under completely positive evolution \cite{Wolf2}.  We show here that this volume signature can witness non-Markovianity if and only the trace of the (generalized) Bloch damping matrix satisfies a simple condition.  For qubits, this condition is strictly stronger than the trace distance condition in Eq.~(\ref{tracecond}).  Hence, the volume provides a weaker signature of non-Markovianity than the trace distance. 

As is well known, a $d$-dimensional quantum system may be represented by a generalized Bloch vector ${\bf x}$ of dimension $d^2-1$, and the corresponding master equation by a generalized Bloch equation of the same form of Eq.~(\ref{bloch}) as for qubits \cite{vanwond,volume}.  This equation is linear, and hence the evolution of the system is given by a linear map of the form
\begin{equation}  \label{map}
{\bf x}(t) = M(t) {\bf x}(0) + {\bf w}(t).  
\end{equation}
Substitution into Eq.~(\ref{bloch}) then yields the equivalent evolution equations
$\dot{M} = DM$, $\dot{\bf w} = {\bf u} + D{\bf w}$,
with initial conditions $M(0)=I$ and ${\bf w}(0)={\bf 0}$.  

The determinant of $M(t)$ determines the Jacobian of the map in Eq.~(\ref{map}), and hence the `Bloch volume' of the states of the system evolves as $V(t)=V_0 \det M(t)$ \cite{volume}.  Now, to first order in $\epsilon$, the above evolution equations imply
$\det M(t + \epsilon) = \det (M  + \epsilon DM) = (\det M) \prod_j (1+\epsilon D_{jj})= (\det M) (1+\epsilon\, \text{tr}[D])$ \cite{trace}. Hence, $\frac{d}{dt} \det M = \text{tr}[D] \det M$, which immediately implies that {\it the Bloch volume can increase at time $t$, thus witnessing non-Markovianity, if and only if the trace of the damping matrix is positive}, i.e., if and only if
\begin{equation} \label{volume}
\text{tr}[D(t)]  > 0 .
\end{equation}
Note that one also immediately obtains the formula
\begin{equation} \label{bvol}
V(t)=V_0 \exp\left( \int_0^t ds\, \text{tr}[D(s)]\right)  
\end{equation}
for the Bloch  volume.

For the case of qubits, condition (\ref{volume}) is clearly always stronger than the corresponding trace distance condition (\ref{tracecond}). As a direct consequence, or alternatively via Eq.~(\ref{damp}), the Bloch volume cannot witness the non-Markovianity of the example in Eqs.~(\ref{ex}) and (\ref{exgamma}). 

More generally, as shown in Appendix~B, 
\begin{equation} \label{volume2}
\text{tr}[D(t)]=-d\sum_k\gamma_k(t) .
\end{equation}
Hence, condition (\ref{volume}) may equivalently be written in terms of the canonical decoherence rates as
$\sum_k \gamma_k(t)  < 0$.
Thus, any measure of non-Markovianity based on the Bloch volume is only sensitive to the {\it sum} of the canonical decoherence rates, both positive and negative. 

\subsection{Entanglement  measures}

The entanglement between two quantum systems cannot increase under local completely positive operations (and/or classical communication), and hence Rivas {\it et al.} have proposed using the increase of any entanglement measure $E$, under local evolution, as a signature of non-Markovianity \cite{Rivas}.  This signature has been  experimentally investigated for the case of concurrence \cite{exp1}. The use of logarithmic negativity \cite{Rivas, neg} and quantum mutual information \cite{inf} have also been proposed.  Here we show that none of these entanglement measures can witness the non-Markovianity of the example in Eqs.~(\ref{ex}) and (\ref{exgamma}).

In particular, writing the system master equation as $\dot \rho_s=\Lambda_t[\rho_s]$, we follow Rivas {\it et al.} and consider the entanglement between the system and an ancilla under the evolution
\begin{equation}
\dot\rho_{sa} = (\Lambda_t\otimes \mathbb{1})[\rho_{sa}] ,
\end{equation}
for an initial maximally entangled state $\rho_{sa}(0)=|\Psi\rangle\langle\Psi|$\cite{Rivas}.   
For the qubit master equation in  Eqs.~(\ref{ex}) and (\ref{exgamma}), we may take $|\Psi\rangle = (|+\rangle\otimes|+\rangle+|-\rangle\otimes|-\rangle)/\sqrt{2}$, with $|\pm\rangle$ denoting the eigenstates of $\sigma_3$. Relative to the basis $\{|+\rangle\otimes|+\rangle,|+\rangle\otimes|-\rangle,|-\rangle\otimes|+\rangle,|-\rangle\otimes|-\rangle\}$, the joint state at time $t$ is then easily found to be
\[
\rho_{sa}(t) =\frac{1}{4}\left(
\begin{array}{cccc}
1+e^{-2t} & 0& 0& 1+e^{-2t}\\
 0& 1-e^{-2t}&0 &0 \\
 0& 0& 1-e^{-2t}&0 \\
1+e^{-2t} &0 &0 & 1+e^{-2t}
\end{array}
\right) .
\]

The corresponding concurrence, logarithmic negativity and quantum mutual information  can all be evaluated analytically, as
\begin{equation}
{\cal C}[\rho_{sa}(t)] =e^{-2t} ,~~~
E_N[\rho_{sa}(t)] =  \log_2 (1+e^{-2t}),
\end{equation}
and
\begin{equation}
I_Q[\rho_{sa}(t)] = 2-H\!\left(\frac{1-e^{-2t}}{2},\frac{1+e^{-2t}}{2}\right)-\frac{1-e^{-2t}}{2} ,
\end{equation}
respectively, where $H(p,1-p)$ denotes the entropy $-p\log_2 p-(1-p)\log_2(1-p)$.
These quantities are manifestly monotonic decreasing with time.  Hence, none of them provide a signature that witnesses the non-Markovianity of the example in Eqs.~(\ref{ex}) and (\ref{exgamma}).

\subsection{Choi-matrix measures}

Any linear map describing the evolution of the density operator of a finite $d$-dimensional quantum system has a corresponding $d^2\times d^2$ Choi matrix $S$, where the evolution is completely positive if and only if $S\geq 0$ \cite{choi1}.  Since Markovianity is equivalent to divisibility into a sequence of infinitesimal completely positive evolutions \cite{Wolf2}, this naturally leads to measures of non-Markovianity based on properties of the Choi matrix \cite{Wolf, Rivas}.  
We show here that the proposed measures are simply related to the canonical measures of non-Markovianity defined in Sec.~III~B, and that they are faithful witnesses, in the sense of always detecting non-Markovianity when it exists.

First, choosing an orthonormal basis $\{|\alpha\rangle\}$ on the Hilbert space, with $\alpha=1,\dots,d$, the action of any linear map $\phi$ on the states of the system can be expanded as
\begin{eqnarray} \nonumber
\phi[\rho] &=& \sum_{\alpha_1,\beta_1} |\alpha_1\rangle\langle\alpha_1| \phi[\rho] |\beta_1\rangle\langle\beta_1|\\ \nonumber
&=& \sum_{\alpha_1,\beta_1,\alpha_2,\beta_2} |\alpha_1\rangle\,\langle\alpha_2|\rho|\beta_2\rangle\,\langle \beta_1|\,\langle\alpha_1|\phi[|\alpha_2\rangle\langle\beta_2|]|\beta_1\rangle\\
&=& \sum_{a,b} S_{ab}\, \tau_a \rho\,\tau_b^\dagger,  \label{choirep1}
\end{eqnarray}
where $a$ and $b$ denote the pairs $(\alpha_1,\alpha_2)$ and $(\beta_1,\beta_2)$, respectively, and
\begin{equation} \label{choi}
S_{ab}:=\langle\alpha_1|\phi[|\alpha_2\rangle\langle\beta_2|]|\beta_1\rangle,~~ \tau_a:= |\alpha_1\rangle\langle\alpha_2|. 
\end{equation}
The $d^2\times d^2$ matrix $S$ is called the Choi matrix corresponding to $\phi$ \cite{choi1,choi2}. Note that the Choi matrix of the identity map, $\mathbb{1}$, is 
\begin{equation} \label{v}
S^\mathbb{1}_{ab} = v_av_b,~~~~~v_a:=\delta_{\alpha_1\alpha_2} .
\end{equation}
It is also straightforward to check that $d^{-1}S$ is the matrix representation of the operator $(\phi\otimes \mathbb{1})[|\Psi\rangle\langle\Psi|]$ in the basis $\{|a\rangle=|\alpha_1\rangle\otimes|\alpha_2\rangle\}$, where $|\Psi\rangle$ is the maximally entangled state $d^{-1/2}\sum_\alpha |\alpha\rangle\otimes|\alpha\rangle$ \cite{choi2}, i.e.,
\begin{equation}
 S_{ab} = d\,\langle a|(\phi\otimes \mathbb{1})[|\Psi\rangle\langle\Psi|]|b\rangle .  \label{choirep2}
\end{equation}
Thus, $S$ is Hermitian if $\phi$ maps Hermitian operators to Hermitian operators, and $\text{tr}[S]=d$ if $\phi$ is trace-preserving \cite{trace}.

As mentioned, $\phi$ is completely positive if and only if $S\geq0$ \cite{choi1,choi2}.  Hence, the infinitesimal map $\phi^\epsilon=\mathbb{1}+\epsilon \Lambda_t$, generated by the master equation $\dot\rho=\Lambda_t[\rho]$, is completely positive, corresponding to Markovian evolution at time $t$, if and only if 
\begin{equation} \label{choimark}
S^\epsilon=S^\mathbb{1}+\epsilon R(t)\geq 0.
\end{equation}  
Here  $R(t)$ denotes the Choi matrix of $\Lambda_t$, obtained by replacing $\phi$ by $\Lambda_t$ in Eq.~(\ref{choi}) or (\ref{choirep2}).  

\subsubsection{Trace-norm measure}

It follows from Eq.~(\ref{choimark}) that (trace preserving) evolution is Markovian at time $t$ if and only the eigenvalues of $S^\epsilon=S^\mathbb{1}+\epsilon R(t)$ are positive and sum to $d$.  Hence, defining the trace-norm rate of change \cite{Rivas}
\begin{equation} \label{gt}
g(t):= \lim_{\epsilon\rightarrow 0^+} \frac{d^{-1}\|S^\mathbb{1}+\epsilon R(t)\|_1-1}{\epsilon} ,
\end{equation}
one has $g(t)\geq 0$, with $g(t)=0$ if and only if the evolution at time $t$ is Markovian.  This led Rivas {\it et al.} to propose 
\begin{equation}
{\cal I} := \int_0^\infty ds \,g(s)
\end{equation}
as a formal measure of non-Markovianity \cite{Rivas}.

It is shown in Appendix~C that the trace-norm measure $g(t)$ has a surprisingly simple interpretation. In particular, it is just the sum of the negative canonical decoherence rates, up to a multiplicative factor:  
\begin{equation} \label{gf}
g(t) = d^{-1}\sum_{k=1}^{d^2-1} \left[ |\gamma_k(t)|-\gamma_k(t)\right] =\frac{2}{d} f(t) ,
\end{equation}
Here $f(t)$ is the canonical measure defined in Eq.~(\ref{ft}).  It follows immediately that
\begin{equation} \label{IF}
{\cal I} = \frac{2}{d}\, F(0,\infty),
\end{equation}
where $F(t,t')$ is the canonical measure of total non-Markovianity defined in Eq.~(\ref{total}). It further follows that $g(t)$ and ${\cal I}$ can always detect non-Markovian evolution \cite{Rivas}, including the example in Sec.~III~D in particular.

The formal connection between the canonical and trace-norm measures follows from a relationship between the decoherence matrix and a projected form of the Choi matrix (Appendix~C).  However, the  canonical approach not only provides a physical interpretation of ${\cal I}$, but also has the advantage of identifying all of the individual decoherence channels and corresponding decoherence rates -- allowing, for example, the definition of the discrete non-Markov index $n(t)$ in Eq.~(\ref{index}).

\subsubsection{Isotropic noise measure}

Wolf {\it et al.} considered the minimal amount of isotropic noise that must be added to a given quantum channel, to allow its simulation by a memoryless master equation \cite{Wolf}.  This may be adapted to obtain a natural measure of the non-Markovianity of a given evolution, at time $t$, corresponding to the smallest amount of isotropic noise, $\epsilon\nu$, that must be added to the infinitesimal evolution $\phi^\epsilon$ to make it completely positive. Here $\nu=\nu(t)$ denotes the rate at which the noise is added at time $t$. As for the trace-norm measure above, $\nu(t)$ turns out to be a simple function of the canonical decoherence rates.

In particular, adding isotropic noise $\epsilon \nu(t)$ is equivalent to mixing the state of the system with the maximally-mixed state, i.e., to the infinitesimal mixing map
\[ \mu^\epsilon[\rho] := (1-\epsilon\nu)\rho + \epsilon \nu \, \text{Tr}[\rho]\, d^{-1}\hat 1 \mg = \rho+\epsilon\nu (d^{-1} \hat1-\rho). \blk \]
The corresponding Choi matrix $N^\epsilon$ follows from  Eqs.~(\ref{choi}) and (\ref{v}) as \cite{Wolf}
\begin{equation} \label{noise}
 N^\epsilon = S^\mathbb{1} +\epsilon \nu(t) \left[d^{-1}I - v v^T\right] ,
\end{equation}
where $I$ denotes the $d^2\times d^2$ identity matrix.

As shown in Appendix~C, the minimum rate of noise $\nu(t)$ that must be added \mg to the master equation  \blk to give completely positive evolution at time $t$ has the explicit simple form
\begin{equation} \label{min}
\nu(t) = d\max_k \{0,-\gamma_1,-\gamma_2,\dots\} = d\max_k \{f_k(t)\},
\end{equation}
where $f_k(t)$ is the canonical decoherence measure for the $k$th decoherence channel, defined in Eq.~(\ref{fk}).  Thus, {\it the minimum noise rate is determined by the value of the most negative canonical decoherence rate}. For the example in Eqs.~(\ref{ex}) and (\ref{exgamma}), $\nu(t)=\tanh t>0$ for all $t>0$.  More generally, Eq.~(\ref{min}) implies that non-Markovianity is always detected by the isotropic noise measure. \mg Note, however, it is not sensitive to {\it all} of the negative decoherence rates, in contrast to the canonical measures $F(t,t')$ and $n(t)$ and the trace-norm measure ${\cal I}$. \blk

\section{Conclusions}

The canonical form for time-local master equations allows a complete characterization of the non-Markovianity of open quantum systems, in terms of a set of uniquely determined canonical decoherence rates. These rates are invariant under time-dependent unitary transformations of the system, and for a given master equation they may be evaluated as the eigenvalues of the decoherence matrix $\mathsf{\mathbf{d}}$ in Eq.~(\ref{dexplicit}), the matrix $R^\perp$ in Eq.~(\ref{rperp}), or the operator ${\cal R}^\perp$ in Eq.~(\ref{rperpop}). 

A positive (negative) canonical decoherence rate corresponds to Markovian (non-Markovian) evolution in the respective decoherence channel. 
This leads naturally to well-defined canonical measures of the degree of non-Markovianity of each channel, and to corresponding measures of the total non-Markovianity of the evolution, as per Eqs.~(\ref{fk})-(\ref{total}).  We have also defined a discrete non-Markov index, in Eq.~(\ref{index}), which characterizes the dimension of the space of non-Markovian evolution.

Signatures of non-Markovianity proposed in the literature can be assessed, both qualitatively and quantitatively, in terms of their relative sensitivities to the presence of negative canonical decoherence rates.  For example, we have shown that signatures based on witnessing an increase in trace distance, Bures distance, Bloch volume, concurrence, logarithmic negativity and quantum mutual information are completely insensitive to the non-Markovianity of a simple qubit evolution. It would be of interest to determine whether there are any measures of distance or entanglement, preferably physically measurable, from  which this evolution cannot successfully hide. 

In contrast, proposed measures based on properties of Choi matrices are always faithful witnesses of non-Markovianity, and we have obtained simple formulas and corresponding physical interpretations of  these measures in terms of the canonical decoherence rates.  Thus, the trace-norm measure of Rivas {\it et al.} is proportional to the sum of the negative decoherence rates, integrated over the time period of the evolution, while the isotropic noise measure of Wolf {\it et al.} is proportional to the most negative decoherence rate.

We have also found simple necessary and sufficient conditions for trace distance and Bloch volume to witness non-Markovian qubit evolution, and that the former provides a strictly stronger signature.  Further, for arbitrary dimensions, we have explicitly determined the Bloch volume as a function of the canonical decoherence rates.

The above results clearly demonstrate the value of the canonical approach in providing a fundamental characterization of non-Markovianity.  It would be of interest to extend this approach to infinite dimensional systems, if possible, and compare with approaches based on divisibility and quantum trajectories in this case. The main technical difficulty appears to be related to non-finite operator traces in this regard.  Finally, it would also be of interest to experimentally characterise the canonical decoherence rates, e.g., via process tomography.

\section*{ACKNOWLEDGMENTS} We thank H. Wiseman and A. Rivas for useful comments. \blk M. H. and L. L. are  supported by the ARC Centre of Excellence CE110001027. E. A. acknowledges financial support by the Royal Society of London, and J. C. by Heriot-Watt University.

\appendix

\section{Derivation of the canonical form for master equations}

A general time-local master equation, such as in Eq.~(\ref{simple}), equates $\dot\rho$ with a linear map $\Lambda_t$ acting on $\rho$, and hence can always be written in the form \cite{Gorini}
\begin{equation} \label{a1}
\dot{\rho}= \Lambda_t[\rho] = \sum_{k}^{}A_k(t)\rho B_k^\dagger(t).
\end{equation} 
The canonical form in Eq.~(\ref{canonical}) essentially then follows via the requirements that $\rho$ remain Hermitian for all time and that the trace of $\rho$ be preserved.

First, for a state space of dimension $d$, define $N:=d^2$ and introduce a complete set of $N$ basis operators $\{G_m;m=0,1,2, \ldots N-1\}$, with the properties
\begin{equation} \label{gorthog}
G_0=\hat 1\big/\big.\!\sqrt{d};\qquad G_m=G_m^\dagger;\qquad \text{Tr}[G_mG_n]=\delta_{mn},
\end{equation}
where $\hat 1$ is the identity operator.
Choosing $n=0$ in the last condition implies that
$\text{Tr}[G_m]=0$ for $ m\ne 0.$

For brevity, we will suppress the time dependence in quantities below, but everything except 
the basis operators $G_m$ may be time dependent. We can expand
$$A_k=\sum_{i}^{}G_ia_{ik},\qquad B_k=\sum_{j}^{}G_jb_{jk}$$
so that Eq.~(\ref{a1}) becomes
$\dot{\rho}=\sum_{i,j}^{}\sum_{k}^{}a_{ik}b_{jk}^*G_i\rho G_j.$
Defining the quantities 
$c_{ij}=\sum_{k}^{}a_{ik}b_{jk}^*$
then yields the unique decomposition \cite{Gorini}
\begin{equation}
\dot{\rho}=\sum_{i,j=0}^{N-1}c_{ij}G_i\rho G_j.
\end{equation}
Using the fact that $\rho$ and hence $\dot{\rho}$ are Hermitian, then
$$\sum_{i,j}^{}c_{ij}G_i\rho G_j=\sum_{i,j}^{}c_{ij}^*G_j\rho G_i=\sum_{i,j}^{}c_{ji}^*G_i\rho G_j,$$
so that
$c_{ij}=c_{ji}^*.$
Thus the $c_{ij}$ are the elements of an $N\times N$ Hermitian matrix. 

Separating out the $i=0$ and $j=0$ terms, the above decomposition for $\dot{\rho}$ reduces to
\begin{eqnarray*}
\dot{\rho}&=&\frac{c_{00}}{d}\rho +\left(\sum_{i=1}^{N-1}\frac{c_{i0}}{\sqrt{d}}G_i\right)\rho + \rho \left(\sum_{j=1}^{N-1}\frac{c_{0j}}{\sqrt{d}}G_j\right)\\
&~&+\sum_{i,j=1}^{N-1}d_{ij}G_i\rho G_j,
\end{eqnarray*}
where  $d_{ij}:=c_{ij}$ for $i,j>0$ are elements of an $(N-1)\times(N-1)$ `decoherence' Hermitian matrix $\mathsf{\mathbf{d}}$. Defining 
\begin{equation}
C:=\half \frac{c_{00}}{d}+\sum_{i}^{}\frac{c_{i0}}{\sqrt{d}}G_i,
\end{equation}
and noting Hermiticity implies that $c_{00}$ is real and $c_{0j}=c_{j0}^*$, this reduces to
\begin{equation}
\dot{\rho}=C\rho+\rho C^\dagger+\sum_{i,j=1}^{N-1}d_{ij}G_i\rho G_j .
\label{canonical2}
\end{equation}

Taking the trace of Eq.~(\ref{canonical2}), and noting that trace preservation implies that $\text{Tr}[\dot{\rho}]=0$, yields
$$C+C^\dagger=-\sum_{i,j=1}^{N-1}d_{ij}G_jG_i.$$
Hence, defining 
$H:=\frac{1}{2}i\hbar (C-C^\dagger)$,
Eq.~(\ref{canonical2}) can be rewritten as
\begin{align}
\nonumber
	\dot{\rho}=&\half\left[(C-C^\dagger)\rho+ \rho(C^\dagger-C)\right.\\
	&\left.+(C+C^\dagger)\rho+\rho(C^\dagger+C)\right]
	+\sum_{i,j=1}^{N-1}d_{ij}G_i\rho G_j\notag\\
	=&-\frac{i}{\hbar}[H,\rho]+\sum_{i,j=1}^{N-1}d_{ij}(t)\left(G_i\rho G_j
	-\half \left\{ G_jG_i,\rho \right\}\right). \label{canonical1}
\end{align}
This is the kind of structure obtained in Theorem 2.2 of~\cite{Gorini}, although because they are considering quantum semigroups, their decoherence matrix $\mathsf{\mathbf{d}}$ is independent of time, as is $H$.  Note that one has the two explicit expressions 
\begin{equation} \label{dexplicit}
d_{ij} =\sum_{k} \text{Tr}[G_iA_k]\,\text{Tr}[G_jB_k^\dagger]=\sum_{m=0}^{d^2-1} \text{Tr}[G_mG_i\Lambda_t[G_m] G_j]
\end{equation}
for the elements of the decoherence matrix, where the first follows directly from the above construction, and the second via the proof of Lemma 2.2 of \cite{Gorini}.

We now observe a crucial feature of this last result (see also \cite{Wolf2,eprint}) that seems not \mg to \blk be widely appreciated, but which enables us to derive the main results of this paper. In particular, we take advantage of the Hermitian nature of the decoherence matrix to write it
in diagonal form,
\begin{equation} \label{diag}
d_{ij}=\sum_{k}^{}U_{ik}\gamma_kU_{jk}^*,
\end{equation}
where the eigenvalues $\gamma_k$ of $\mathsf{\mathbf{d}}$ are real, but not necessarily positive at all times, and the $U_{ik}$ are unitary $(N-1)\times(N-1)$ matrices formed by the corresponding eigenvectors of $\mathsf{\mathbf{d}}$, with
$\sum_{k}^{}U_{ik}U_{jk}^*=\delta_{ij}$.
Defining 
the \emph{time dependent} operators
\begin{equation}
L_k(t):=\sum_{i=1}^{N-1}U_{ik}(t)G_i,
\end{equation}
and restoring explicit time-dependence, we finally obtain the {\it canonical} form in Eqs.~(\ref{canonical}) and (\ref{orthog})of the text via Eqs.~(\ref{gorthog}) and \eqref{canonical1}. Noting that only the traceless part of $H(t)$ can contribute to the commutator in Eq.~(\ref{canonical}), one may further assume $\text{Tr}[H(t)]=0$ without any loss of generality.

The decoherence rates $\gamma_k(t)$ are always uniquely defined, as a consequence of the diagonal decomposition in Eq.~(\ref{diag}). Similarly, the decoherence operators $L_k(t)$ are unique up to unitary transformations that preserve any degenerate subspaces of the decoherence matrix $\mathsf{\mathbf{d}}$.  For the case of a nondegenerate decoherence matrix, this implies uniqueness of $L_k(t)$ up to multiplication by a trivial phase factor, and hence that the canonical form of the master equation in Eq.~(\ref{canonical}) is fully fixed in this case.

It is  of interest to note that under any unitary transformation $\rho\rightarrow V(t)\rho V(t)^\dagger$, e.g. to an `interaction' picture, the $\gamma_k(t)$ are invariant (whereas both $H(t)$ and the $L_k(t)$ will typically change), since any such transformation corresponds to an orthogonal transformation of the decoherence matrix. This has the consequence that non-Markovianity can be defined independently of the operator $H$. 

It is also worth noting that is not in general possible to transform to a picture in which the $L_k$ are time independent. 
This is because although a suitable (time dependent) orthogonal transformation of the $L_k(t)$ always exists in the linear space of traceless operators, that preserves the relations in Eq.~(\ref{orthog}), such a transformation will not always correspond to some  
unitary transformation on a $d \times d$ density matrix (unless $d=2$).

\section{Bloch volume and the decoherence matrix}

In Sec.~IV~B it was shown that the condition for the Bloch volume to increase at time $t$ is that the trace of the damping matrix $D(t)$ is positive.  To show this is equivalent to the trace of the decoherence matrix $\mathsf{\mathbf{d}}$ being negative, as per Eq.~(\ref{volume2}), define the generalised Bloch vector ${\bf x}$ by $x_m:=\text{Tr}[\rho G_m]$, $m=1,2,\dots,d^2-1$, with $G_m$ as per Eq.~(\ref{gorthog}) \cite{volume,vanwond}. It follows from $\rho=\sum_{n\geq 0} \text{Tr}[\rho G_n]\,G_n$  that
\[
\dot x_m = \text{Tr}[\Lambda_t[\rho]G_m] = \sum_{n\geq0} \text{Tr}[\rho G_n]\,\text{Tr}[\Lambda_t[G_n]G_m]. \]
Hence, the damping matrix coefficients follow as  $D_{mn}=\text{Tr}[\Lambda_t[G_n]G_m]$ for $m,n\geq1$, and so, using the canonical form of the master equation,
\begin{align*}
D_{mm}&=-\frac{i}{\hbar}\text{Tr}[[H,G_m]G_m] \\
&~+\sum_k\gamma_k\text{Tr}[L_k^\dagger G_mL_kG_m-L_k^\dagger L_k (G_m)^2].
\end{align*}
The first term trivially vanishes.  Further, the completeness property $\sum_{m\geq0} G_mXG_m = \text{Tr}[X] \hat 1$ \cite{Gorini} implies that $\sum_{m\geq1} G_mXG_m=\text{Tr}[X] \hat 1-d^{-1}X$, and hence, choosing $X=L_k$ and $X=\hat 1$ as appropriate, the trace of the damping matrix evaluates to \cite{trace}
\[ \text{tr}[D]=\sum_{m\geq1}D_{mm} = -d\sum_k\gamma_k \text{Tr}[L_k^\dagger L_k]. \]
Finally, $\text{Tr}[L_k^\dagger L_k]=1$ from Eq.~(\ref{orthog}), and the claimed equivalence between Eqs.~(\ref{volume}) and (\ref{volume2}) follows as desired.

Note that the sum of the canonical decoherence rates in Eq.~(\ref{volume2})  only requires evaluation of the trace of the decoherence matrix $\mathsf{\mathbf{d}}$ in Eq.~(\ref{dexplicit}).  This implies that no diagonalization of $\mathsf{\mathbf{d}}$ (nor writing the master equation in Bloch form) is needed to determine the Bloch volume $V(t)$ in Eq.~(\ref{bvol}).

\section{Choi matrix and canonical decoherence rates}

The Markovianity condition (\ref{choimark}) can be rewritten, using Eq.~(\ref{v}), as
$S^\epsilon = v v^T + \epsilon R \geq 0$.
This clearly holds (for infinitesimal $\epsilon$) if and only if $w^TR(t)w\geq 0$ for all $w$ orthogonal to $v$, i.e., if and only if
\begin{equation} \label{rperp}
 R^\perp:=(I-P) R(I-P) \geq 0,
\end{equation}
where $P=d^{-1}vv^T$ denotes the projection on to the direction of $v$.  Further, recalling $\text{tr}[S^\epsilon]=d$ \cite{trace}, it also follows from Eq.~(\ref{choimark}) that
\begin{eqnarray} \nonumber
0 &=& \text{tr}[R] = \text{tr}[PRP] + \text{tr}[(I-P)R(I-P)]\\
&=& d^{-1} v^TRv+ \text{tr}[ R^\perp]. \label{alpha}
\end{eqnarray}
Now, the eigenvalues of $vv^T$ are trivially $d,0,0,\dots$.  Choosing the corresponding eigenvectors as $d^{-1/2}v$, $w_1$, $w_2,\dots$, where $w_k=(1-P)w_k$ is the $k$th eigenvector of $R^\perp$,  the  eigenvalues of $S^\epsilon=v v^T + \epsilon R$ follow from standard perturbation theory as $d+\epsilon d^{-1} v^TRv = d-\epsilon\text{tr}[R^\perp]$ and  
\[ \epsilon w_k^TRw_k =\epsilon w_k^T(1-P)R(1-P)w_k =\epsilon w_k^TR^\perp w_k = \epsilon\,r_k, \]
to first order in $\epsilon$, where $r_1,r_2,\dots,r_{d^2-1}$ denote the eigenvalues of $R^\perp$.  It follows immediately that the quantity $g(t)$ defined in Eq.~(\ref{gt}) can be written as
\begin{equation} \label{gr}
g(t) = d^{-1} \left[- \text{tr}[R^\perp] + \sum_k |r_k|\right] = d^{-1} \sum_k \left[ |r_k|-r_k\right]. 
\end{equation}

We now show that $r_k=\gamma_k$.  Expanding $R=[(P+(I-P)]R[(P+(I-P)]$ and using Eq.~(\ref{alpha}) gives
\[ R=\alpha vv^T + vw^\dagger + wv^T +R^\perp, \]
where $w$ is  orthogonal to $v$ and $\alpha=-d^{-1}\text{tr}[R^\perp]$. Recalling $R$ is defined to be the Choi matrix of $\Lambda_t$, it immediately follows from Eq.~(\ref{choirep1}) that the master equation can be written as
\begin{eqnarray*}
\dot\rho&=&\Lambda_t[\rho] = \sum_{a,b} R_{a,b}(t)\,\tau_a \rho\,\tau_b^\dagger\\
&=&\rho (\half\alpha+W^\dagger) +(\half\alpha+W)\rho +\sum_k r_k W_k\rho W_k^\dagger,
\end{eqnarray*}
in terms of $W:=\sum_a w_a\tau_a$, $W_k:=\sum_a w_{k,a}\tau_a$ and the eigenvalue decomposition $R^\perp=\sum_k r_k w_kw_k^\dagger$, where we have used Eq.~(\ref{v}) to write $\sum_a v_a\tau_a=\hat 1$. Decomposing $\half\alpha+W=J-i\hbar K$ into Hermitian and antiHermitian parts, and applying the trace-preserving condition $\text{Tr}[\Lambda_t[\rho]]=\text{Tr}[\rho]$, yields $2J=-\sum_k r_k W_k^\dagger W_k$.  Thus, restoring explicit time dependence,
\begin{eqnarray}	\nonumber
	\dot\rho
	&=&-\frac{i}{\hbar} [K(t),\rho] 
	+\sum_{k=1}^{d^2-1}r_k(t)\left[W_k(t)\rho W_k^\dagger(t) \right. \\
&~&~~~~~~~~~~~~~~~~~~~~~~~\left. -\half \left\{ W_k^\dagger(t) W_k(t),\rho \right\} \right]. \label{canonicalchoi}
\end{eqnarray}
Note also, using $v\cdot w_k=0$ and $w_k\cdot w_l=\delta_{kl}$, that
\begin{equation} \label{orthogchoi}
\text{Tr}[W_k(t)] = 0,~~~~~\text{Tr}[W_j^\dagger(t)\, W_k(t)] =\delta_{jk} .
\end{equation}
Comparison with Eqs.~(\ref{canonical}) and (\ref{orthog}) shows that Eq.~(\ref{canonicalchoi}) is in canonical form.  Hence, since the canonical decoherence rates are uniquely determined by the canonical form (Appendix~A), we have
\begin{equation} \label{rgamma}
\gamma_k(t) = r_k(t)
\end{equation}
for some suitable ordering of the eigenvalues of $R^\perp$.  

Note that while $R^\perp$ has the same eigenvalues as the decoherence matrix $\mathsf{\mathbf{d}}$ in Eq.~(\ref{dexplicit}), allowing the canonical decoherence rates to be calculated in two different ways, these matrices are not equal, but are related by a (nontrivial) unitary transformation.  However, the corresponding defining conditions for Markovianity, $\mathsf{\mathbf{d}}\geq 0$ and $R^\perp\geq 0$, are of course equivalent, and thus the above result explicitly demonstrates the equivalence of Markovian evolution with completely positive infinitesimal maps. 

The formulas in Eqs.~(\ref{gf}) and (\ref{IF}) of the main text, for the trace-norm measures $g(t)$ and ${\cal I}$ proposed by Rivas {\it et al.} \cite{Rivas}, follow immediately from Eqs.~(\ref{gr}) and (\ref{rgamma}).

Moreover, from Eqs.~(\ref{choimark}) and (\ref{noise}), the infinitesimal evolution generated by the addition of isotropic noise at rate $\nu(t)$ to $\Lambda_t$ has the corresponding Choi matrix
\[ T^\epsilon = S^{\mathbbm{1}} +\epsilon\left[R +\nu(d^{-1}I -vv^T)\right] . \]
Similarly as for Eq.~(\ref{rperp}), this corresponds to completely positive evolution if and only if 
\begin{eqnarray*}
0&\leq& (I-P)\left[R +\nu(d^{-1}I -vv^T)\right](I-P)\\
&=& R^\perp + \nu d^{-1} (I-P)\\
&=& \sum_k[r_k+ d^{-1}\nu] w_k w_k^T .
\end{eqnarray*}
 Since $\nu(t)\geq 0$ is, by definition, the smallest noise rate for which the evolution is completely positive (Sec.~IV~D.2), one immediately has
\begin{equation}
\nu(t) = d\max_k \{0,-r_1,-r_2,\dots\} ,
\end{equation}
and Eq.~(\ref{min}) follows via Eq.~(\ref{rgamma}).

Finally, we remark that the projected Choi matrix $R^\perp$ in Eq.~(\ref{rperp}) can alternatively be written as $R^\perp_{ab}=\langle a|{\cal R}^\perp|b\rangle$, analogously to  Eq.~(\ref{choirep2}), with
\begin{equation} \label{rperpop}
{\cal R}^\perp:=d\left(\hat1-|\Psi\rangle\langle\Psi|\right)(\Lambda_t\otimes\mathbbm{1})[|\Psi\rangle\langle\Psi|]\left(\hat1-|\Psi\rangle\langle\Psi|\right).
\end{equation}
Thus the canonical decoherence rates can also be calculated as the eigenvalues of this operator, for any maximally entangled state $|\Psi\rangle$.


\begin{thebibliography}{99}
\bibitem{trace} We follow the practice of using $\text{Tr}[~]$ and $\text{tr}[~]$ to denote the operator trace and the matrix trace, respectively.
\end{thebibliography}
\end{document}